\documentclass [11pt]{article}
\usepackage{amstex}
\begin{document}
\title{{\bf Maximally entangled states and the Bell inequality}\thanks
{An almost identical version of this paper was originally published in:
J.L. Cereceda, Phys. Lett. A {\bf 212}, 123 (1996).}}
\author{Jos\'{e} L. Cereceda \\
{\em C/Alto del Le\'{o}n 8, 4A, 28038 Madrid, Spain}}
\date{December 4, 1998}
\maketitle
\begin{abstract} Kar's recent proof showing that a maximally entangled state of two
spin-1/2 particles gives the largest violation of a Bell inequality is extended to
$N$ spin-1/2 particles \( (N \geq 3) \). In particular, it is shown that all the states
yielding a direct contradiction with the assumption of local realism do generally
consist of a superposition of maximally entangled states.
\end{abstract}
\vspace{5mm}

Recently, Kar (see Ref. \cite{Kar}, and references therein) has shown that a maximally
entangled states of two spin-1/2 particles not only gives a maximum violation of the CHSH
inequality \cite{CHSH} but also gives the largest violation attainable for any pairs of
four spin observables, these pairs being noncommuting for both systems. To prove this,
Kar made use of an elegant (and powerful) technique based on the determination of the
eigenvectors and eigenvalues of the associated Bell operator \cite{BMR}. In this Letter
we would like to extend these results to the case in which $N$ spin-1/2 particles
\( (N \geq 3) \) are considered. We will show that the most general {\it N}-particle state
giving the largest violation of a Bell inequality does consist of a superposition of
maximally entangled states. As expected, those states giving {\em maximal} departure
from classical expectations correspond to the class of states introduced by Greenberger,
Horne, and Zeilinger in proving Bell's theorem without using inequalities \cite{GHZ,GHSZ}.

In order to look for a violation of local realism when dealing with {\em N} spin-1/2
particles it is necessary to consider correlation functions involving measurements on each
of the particles. A suitable generalisation of the CHSH inequality to an {\em N}-measurement
scheme was obtained by Hardy \cite{Hardy}. Hardy's inequalities can be written in the
form
\newcommand{\B}{$B_{{\rm H}}$}
\begin{equation}
-2 \leq \left\langle B_{{\rm H}} \right\rangle \leq 2 \: ,
\end{equation}
where \( \left\langle B_{{\rm H}} \right\rangle \) denotes the expectation value of some
Hermitian operator \B\ (the so-called Bell operator \cite{BMR}) acting on the
$2^{N}$-dimensional tensor product space associated with {\em N} spin-1/2 particles. For
concreteness, and for purposes of comparison with Ref. \cite{Kar}, in what follows we
concentrate on the case $N=3$. Later on, we shall consider the case of arbitrary {\em N}.
For three spin-1/2 particles the representative Bell operator is \cite{Hardy}
\begin{eqnarray}
B_{{\rm H}}\!\!\!& = & \!\!\! \sigma \left({\bf n}_{1} \right)\sigma \left({\bf n}'_{2}\right)
                 \sigma \left({\bf n}'_{3} \right) +
                 \sigma \left({\bf n}'_{1} \right) \sigma \left({\bf n}_{2} \right)
                 \sigma \left({\bf n}'_{3} \right) \nonumber  \\
     &   & \!\! + \,  \sigma \left({\bf n}'_{1} \right) \sigma \left({\bf n}'_{2} \right)
                 \sigma \left({\bf n}_{3} \right) -
                 \sigma \left({\bf n}_{1} \right) \sigma \left({\bf n}_{2} \right)
                 \sigma \left({\bf n}_{3} \right),
\end{eqnarray}
where the observable operator $\sigma \left({\bf n}_{i} \right)$ $\left( \sigma 
\left({\bf n}'_{i} \right) \right)$ corresponds to a spin measurement on particle
{\em i} $(i=1,2,3)$ along the direction ${\bf n}_{i}$ $\left( {\bf n}'_{i} \right)$. 
The square of this Bell operator is given by
\begin{eqnarray}
B_{{\rm H}}^{2} \: = \: 4I \!\! \!& - & \!\!\! \left[ \sigma \left({\bf n}_{1} \right),
\,  \sigma \left({\bf n}'_{1}
\right) \right] \left[ \sigma \left({\bf n}_{2} \right), \, \sigma \left({\bf n}'_{2}
\right) \right] \nonumber \\
 & - &  \!\!\!\left[ \sigma \left({\bf n}_{2} \right), \, \sigma \left({\bf n}'_{2}
\right) \right] \left[ \sigma \left({\bf n}_{3} \right), \, \sigma \left({\bf n}'_{3}
\right) \right] \nonumber \\
 & - &  \!\!\!\left[ \sigma \left({\bf n}_{1} \right), \, \sigma \left({\bf n}'_{1}
\right) \right] \left[ \sigma \left({\bf n}_{3} \right), \, \sigma \left({\bf n}'_{3}
\right) \right] .
\end{eqnarray}
Now, replacing each commutator \( \left[ \sigma \left({\bf n}_{i} \right), \, \sigma
\left({\bf n}'_{i} \right) \right] \) by its value\footnote{This expression for the
commutator actually differs from that of Ref. \cite{Kar} by a factor $-1$. This is because,
following the usual convention, we take here the angle $\theta_{i}$ as the negative of
that used in Ref. \cite{Kar}.} $2 \,{\rm i}\,\sin \theta_{i}\: \sigma_{\bot i}$, we get
\newcommand{\SB}{$B_{{\rm H}}^{2}$}
\begin{eqnarray}
B_{{\rm H}}^{2}\: = \:4I \!\!\!\!& + &\! \!\! 4\,\sin\theta_{1}\,\sin\theta_{2}\:
\sigma_{\bot 1}\, \sigma_{\bot 2}  \nonumber \\
 & + & \!\!\! 4\,\sin\theta_{2}\,\sin\theta_{3}\:\sigma_{\bot 2}\,\sigma_{\bot 3}
\nonumber  \\
 & + & \!\!\!  4\,\sin\theta_{1}\,\sin\theta_{3}\:\sigma_{\bot 1}\,\sigma_{\bot 3} \,,
\end{eqnarray}
where $\theta_{i}$ is the angle included between ${\bf n}_{i}$ and ${\bf n}'_{i}$, and
$\sigma_{\bot i}$ is the spin operator for particle {\em i} along the direction
perpendicular to both ${\bf n}_{i}$ and ${\bf n}'_{i}$. From expression (4) we can see
at once that the largest eigenvalue of \SB\ is
\begin{equation}
\mu_{l}=4 \left( 1+ \left| \sin\theta_{1}\, \sin\theta_{2} \right|
                  + \left| \sin\theta_{2}\, \sin\theta_{3} \right|
                  + \left| \sin\theta_{1}\, \sin\theta_{3} \right| \, \right),
\end{equation}
which attains a maximum value of $\mu_{{\rm max}}=16$. From (4) it is also apparent that
to every eigenvalue of \SB\ there corresponds a pair of (degenerate) eigenvectors, namely,
\mbox{$\left| \,\sigma_{\bot 1},\, \sigma_{\bot 2},\, \sigma_{\bot 3} \right\rangle$} and
\mbox{$\left|\, -\sigma_{\bot 1},\, -\sigma_{\bot 2},\, -\sigma_{\bot 3} \right\rangle$},
where $\left| \, \sigma_{\bot i} \right\rangle$ $\left(\left| \, -\sigma_{\bot i} \right
\rangle \right)$ is the eigenvector of $\sigma_{\bot i}$ with eigenvalue $\sigma_{\bot i}
=\pm1$ $\left( -\sigma_{\bot i}=\mp1 \right)$. In particular, the eigenvectors corresponding
to the largest eigenvalue (5) are (in an obvious notation): \mbox{$\left| \uparrow,\, \uparrow,
\, \uparrow \, \right\rangle$} and \mbox{$\left| \downarrow,\, \downarrow,
\, \downarrow \, \right\rangle$} for ${\rm sgn}\left( \sin\theta_{1} \right)=
{\rm sgn}\left( \sin\theta_{2} \right)$
\linebreak
$={\rm sgn}\left( \sin\theta_{3} \right)$;
\mbox{$\left| \uparrow,\, \uparrow,
\, \downarrow \, \right\rangle$} and \mbox{$\left| \downarrow,\, \downarrow,
\, \uparrow \, \right\rangle$} for ${\rm sgn}\left( \sin\theta_{1} \right)=
{\rm sgn}\left( \sin\theta_{2} \right)$
\linebreak
$\neq {\rm sgn}\left( \sin\theta_{3} \right)$;
\mbox{$\left| \uparrow,\, \downarrow,
\, \downarrow \, \right\rangle$} and \mbox{$\left| \downarrow,\, \uparrow,
\, \uparrow \, \right\rangle$} for ${\rm sgn}\left( \sin\theta_{1} \right) \neq
{\rm sgn}\left( \sin\theta_{2} \right)$
\linebreak
$ = {\rm sgn}\left( \sin\theta_{3} \right)$; and, finally,
\mbox{$\left| \uparrow,\, \downarrow,
\, \uparrow \, \right\rangle$} and \mbox{$\left| \downarrow,\, \uparrow,
\, \downarrow \, \right\rangle$} for ${\rm sgn}\left( \sin\theta_{1} \right)$
\linebreak
$ = {\rm sgn}\left( \sin\theta_{3} \right) \neq {\rm sgn}\left( \sin\theta_{2} \right)$.
On the other hand, it can be easily seen that the minimum possible eigenvalue for
\SB\ is zero. So, for example, $\left| \uparrow,\, \uparrow,\, \downarrow \,
\right\rangle$ is an eigenvector of \SB\ with zero eigenvalue whenever $\theta_{1}
=\theta_{2}=\theta_{3}=\pi /2$. Of course, the operator \SB\ cannot have negative
eigenvalues because this would imply the (Hermitian) operator \B\ has a complex
spectrum.

As every eigenvalue for \SB\ must lie in the interval $[0,16]$ it follows that the
eigenvalues for \B\ are necessarily restricted to lie in the interval $[-4,4]$.
Consequently, inequality (1) for the Bell operator (2) will be violated for those
eigenvectors of \B\ with eigenvalues $\lambda$ fulfilling $2<|\lambda| \leq 4$. Note
that the maximum amount of violation of Hardy's inequality predicted by quantum mechanics
is by a factor of 2, instead of the factor $\sqrt{2}$ achieved in the CHSH inequality.
Moreover, it is worth pointing out that the results mentioned in the present paragraph
remain valid for an arbitrary number of particles. This is so because Hardy-type inequalities
involve only four correlation functions regardless of the number $N$ of measurements
considered \cite{Hardy}. This fact will be used later in considering the $N$-particle
case.

In view of Eq. (5), the largest eigenvalue of \B\ will be
\begin{equation}
\left| \lambda_{l} \right| = 2 \left( 1+ \left| \sin\theta_{1}\, \sin\theta_{2} \right|
                  + \left| \sin\theta_{2}\, \sin\theta_{3} \right|
                  + \left| \sin\theta_{1}\, \sin\theta_{3} \right| \, \right)^{1/2}.
\end{equation}
Now, as a product state cannot give rise to violations of local realism (see below), it
follows that an eigenvector of \B\ with eigenvalue $\lambda$ such that \mbox{$2<|\lambda
|\leq 4$} must necessarily consist of an entanglement of the two degenerate eigenvectors
of \SB\ with eigenvalue $\lambda^{2}$ \cite{BMR},
\begin{equation}
\left| \Psi \right\rangle = \alpha \left|z_{1},\, z_{2},\, z_{3} \right\rangle +
           \beta \, e^{{\rm i} \phi} \left| \, -z_{1},\, -z_{2},\, -z_{3} \right\rangle,
\end{equation}
where, for simplicity, we restrict ourselves to directions ${\bf n}_{i}$ and
${\bf n}'_{i}$ lying in the {\em x}--{\em y} plane (these directions being specified
by the azimuthal angles $\phi_{i}$ and $\phi'_{i}$, respectively), so that $\left|
z_{i} \right\rangle$ $\left(\, \left| \, -z_{i} \right\rangle \right)$ designates the
eigenvector of the spin operator along the $z$-axis for particle $i$, with eigenvalue
$z_{i}=\pm1$ $(-z_{i}=\mp1)$. It turns out (see Eq. (8) below) that the
relative signs of $z_{1}$, $z_{2}$, and $z_{3}$ in Eq. (7) are uniquely determined by
$\lambda$ (for fixed values of $\theta_{1}$, $\theta_{2}$, and $\theta_{3}$.) Likewise,
the real coefficients $\alpha$ and $\beta$ (which are assumed to satisfy the
normalisation condition $\alpha^{2} + \beta^{2} = 1$), as well as the phase factor
$\phi$ will depend on the eigenvalue $\lambda$. We will now show that for the case in which
the eigenvector (7) is associated with the largest eigenvalue (6), these coefficients
must fulfil the condition $|\alpha|=|\beta|=1/{\sqrt{2}}$. In other words, the largest
possible violation occurs for maximally entangled states. This can be seen by directly
evaluating the expectation value for the state vector (7). This is given by
\begin{eqnarray}
\left \langle \Psi \left| B_{{\rm H}} \right| \Psi \right\rangle \, = \,
2\, \alpha \beta \; [ & \!\!\!\!\!\!\!\!\cos \left( \phi - z_{1}\phi_{1} - z_{2}
\phi'_{2} - z_{3} \phi'_{3} \right)  \nonumber \\
  + & \!\!\!\!\!\!\!\! \cos \left( \phi - z_{1}\phi'_{1} - z_{2}\phi_{2} - z_{3}\phi'_{3}
\right)  \nonumber  \\
  + & \!\!\!\!\!\!\!\! \cos \left( \phi - z_{1}\phi'_{1} - z_{2}\phi'_{2} - z_{3}\phi_{3}
\right) \nonumber  \\
  - & \!\!\!\!\! \left.  \cos \left( \phi - z_{1}\phi_{1} - z_{2}\phi_{2} - z_{3}
\phi_{3} \right) \, \right].
\end{eqnarray}
As the product $\alpha \beta$ factorises out in this expression, it is clear that in order
for (8) to reach its largest value (6) it is necessary that the absolute value of
$\alpha\beta$ be a maximum, i.e., $|\alpha|=|\beta|=1/{\sqrt{2}}$. Although it is apparent
from (8) that this must be the case for $\left| \lambda_{l} \right| = 4$, it might seem to
be worthwhile checking explicitly the above statement for the case that $\left| \lambda_{l}
\right| < 4$ by considering a concrete example. So, let us take the values $\phi_{1}=
\phi_{2}=\phi_{3}=0$, $\phi'_{1}=\phi'_{2}=\pi /2$, and $\phi'_{3}=\pi /4$, so that
$\theta_{1}=\theta_{2}=\pi /2$ and $\theta_{3}=\pi /4$. For these values we find that
(see Eq. (6)) $\left| \lambda_{l} \right| = 2 \,( 2+{\sqrt{2}} \,)^{1/2}$. On the other hand,
the absolute value of $\left \langle \Psi \left| B_{{\rm H}} \right| \Psi \right\rangle$
is found to be, $4 \, |\alpha\beta| \: |\, 2^{-1/2} \sin\phi - (1 + 2^{-1/2})
\cos\phi \,|$, where we have put $z_{1}=z_{2}=z_{3}=+1$ in Eq. (8) since, for the above
values for $\theta_{i}$, we have ${\rm sgn}\left( \sin\theta_{1} \right)=
{\rm sgn}\left( \sin\theta_{2} \right) = {\rm sgn}\left( \sin\theta_{3} \right) $ (of
course the reasoning remains essentially unchanged if we instead choose $z_{1}=z_{2}=
z_{3}=-1$; the important point is that ${\rm sgn} \,z_{1}= {\rm sgn} \,z_{2}=
{\rm sgn} \,z_{3}$). Therefore, for some $\alpha$, $\beta$, and $\phi$, we must have
\begin{equation}
2 \, \left|\alpha\beta \right| \, \left|\, 2^{-1/2} \sin\phi - (1+ 2^{-1/2} )
\cos\phi \, \right| = \left( 2+ \sqrt{2} \, \right)^{1/2} \, .
\end{equation}
The only values for which this equality holds are $|\alpha|=|\beta|=1/\sqrt{2}$, and
$\phi = -\pi /8 + n\pi$, where $n=0,\pm1,\pm2,\ldots$ . This follows at once from the
fact that the function $h(\phi) = \mbox{ $|\,2^{-1/2} \sin\phi - (1+2^{-1/2}) \cos
\phi \,|$ }$ reaches its maximum value for $\phi = -\pi /8 + n\pi$, and this value is
precisely $(2+ \sqrt{2} )^{1/2}$.

From the preceding example it is obvious that, if expression (8) is to be {\em equal}
to the largest eigenvalue (6), the phase factor $\phi$ must be a suitable function
depending only on the angles $\phi_{i}$ and $\phi'_{i}$ (or, equivalently, on the
angles $\phi_{i}$ and $\theta_{i}$, as $\theta_{i}=\phi'_{i}\, - \,\phi_{i}$). This is so
because for the largest eigenvalue case the condition $2\, |\alpha\beta|=1$ is always
met, and then Eq. (8) involves only the variables $\phi$, $\phi_{i}$, and $\phi'_{i}$.
This dependence can be easily obtained for the special (and important) case where
$\theta_{i}= \pi /2$ $(i=1,2,3)$. In this case the eigenvalue (6) attains its maximum
value 4, and then the following four equalities should be simultaneously fulfilled
\begin{gather}
\cos \left( \phi - \phi_{1} - \phi'_{2} - \phi'_{3} \right) \, = \,
\pm1 \, , \tag{10a} \\
\cos \left( \phi - \phi'_{1} - \phi_{2} - \phi'_{3} \right) \,  =  \,
\pm1 \, , \tag{10b} \\
\cos \left( \phi - \phi'_{1} - \phi'_{2} - \phi_{3} \right) \, = \, 
\pm1 \, , \tag{10c} \\
\! \cos \left( \phi - \phi_{1} - \phi_{2} - \phi_{3} \right)  \, = \,  \mp1 \, , 
\tag{10d}
\setcounter{equation}{11}
\end{gather}
where we have put $z_{1}=z_{2}=z_{3}=+1$ in (8) since, as before, ${\rm sgn}\left( 
\sin\theta_{1} \right)= {\rm sgn}\left( \sin\theta_{2} \right) = {\rm sgn}\left( 
\sin\theta_{3} \right)$. So, recalling the relationship $\phi'_{i}=\phi_{i}+\pi /2$,
it is a trivial matter to see that equalities (10a)--(10d) can be matched provided that
$\phi=\phi_{1}+\phi_{2}+\phi_{3}+n\pi$, $n=0,\pm1,\pm2,\ldots$ . It thus follows
that the state vector (see Eq. (7))
\begin{equation}
\left| \Psi \right\rangle = {\frac{1}{\sqrt{2}}} \, ( \,\left| \, +,+,+ \right\rangle
\pm e^{{\rm i}(\phi_{1}+\phi_{2}+\phi_{3})} \left| \, -,-,- \right\rangle ) \, ,
\end{equation}
will be an eigenvector of the Bell operator
\begin{eqnarray}
B_{{\rm H}} \!\!\!\!& = & \!\!\!\! \alpha\, \sigma(\phi_{1})\, \sigma(\phi_{2}
+\pi /2)\, \sigma(\phi_{3}
+\pi /2) +  \beta\, \sigma(\phi_{1}+\pi /2)\, \sigma(\phi_{2})\, \sigma(\phi_{3}
+\pi /2)  \nonumber  \\
 &  & + \, \gamma\, \sigma(\phi_{1}+\pi /2)\, \sigma(\phi_{2}+\pi /2)\, \sigma(\phi_{3})
+  \delta\, \sigma(\phi_{1})\, \sigma(\phi_{2})\, \sigma(\phi_{3}) \, , \nonumber \\
 &  &  
\end{eqnarray}
with eigenvalue $\mp4$, whenever $\alpha=\beta=\gamma=-\delta=+1$. More generally,
it can be shown that any state of the form
\begin{equation}
\left| \Psi \right\rangle = {\frac{1}{\sqrt{2}}} \, (\, \left| \,
z_{1},\, z_{2},\, z_{3} \right\rangle
\pm e^{{\rm i}(z_{1}\phi_{1}+z_{2}\phi_{2}+z_{3}\phi_{3})} \left|
\, -z_{1},\, -z_{2},\, -z_{3} \right\rangle ) \, ,
\end{equation}
is an eigenvector of the Bell operator (12) with eigenvalue $+4$ or $-4$ for a suitable
choice of the sign factors $\alpha$, $\beta$, $\gamma$, and $\delta$ (provided, in any
case, that $\alpha\beta\gamma\delta=-1$). In this way, it is clear that the product of the
quantum expectation values for the operators $T_{1}=\sigma(\phi_{1})\, \sigma(\phi_{2}
+\pi /2)\, \sigma(\phi_{3}+\pi /2)$, $T_{2}=\sigma(\phi_{1}+\pi /2)\, \sigma(\phi_{2})\,
\sigma(\phi_{3} +\pi /2)$, $T_{3}=\sigma(\phi_{1}+\pi /2)\, \sigma(\phi_{2}+\pi /2)\,
\sigma(\phi_{3})$, and $T_{4}=\sigma(\phi_{1})\, \sigma(\phi_{2})\, \sigma(\phi_{3})$,
must be equal to $-1$ when evaluated for any of the states (13). Indeed, the set of
vectors (13) forms a basis set (with a total of eight linearly independent vectors)
which simultaneously diagonalises the four (commuting) operators $T_{1}$, $T_{2}$,
$T_{3}$, and $T_{4}$ \cite{Cereceda}. As a result, the value $-1$ for the above product
of expectation values actually arises from the fact that $T_{1}\,T_{2}\,T_{3}\,T_{4}
=-I$. That these quantum predictions for the expectation values indeed lead to a direct
contradiction with the assumption of local realism constitutes the theorem of
Greenberger, Horne, and Zeilinger \cite{GHZ,GHSZ} (see also Refs. {\cite{Cereceda,
Mermin}.) In fact, it can be easily shown \cite{Hardy} that a maximum violation of
inequality (1) always entails a nonlocality contradiction of the GHZ type.

Before analysing the $N$-particle case, it is worth noting an implication of Eq. (8).
Indeed, from that equation it follows that, for fixed $\phi$, the eigenvalue $\lambda$
does not determine the individual values of $\alpha$ and $\beta$ but, rather, the value
of their product $\alpha\beta$. This means that, whenever $|\alpha| \neq |\beta|$, if the
state vector $\left| \Psi_{\alpha\beta} \right\rangle = \alpha \left|z_{1},\, z_{2},
\, z_{3} \right \rangle + \beta \, e^{{\rm i} \phi} \left| \, -z_{1},\, -z_{2},
\, -z_{3} \right\rangle$ happens to be an eigenvector of (2) with associated eigenvalue
$\lambda$, the same is true for the (linearly independent) vector $\left| \Psi_{\beta
\alpha} \right\rangle = \beta \left|z_{1},\, z_{2}, \, z_{3} \right \rangle +
\alpha \, e^{{\rm i} \phi} \left| \, -z_{1},\, -z_{2}, \, -z_{3} \right\rangle$. This
degeneracy is due to the very structure of the state vector (7), and will be called
here a {\em trivial} degeneracy. Notice that the {\em trivial} degeneracy is removed
when $\lambda$ corresponds to the largest eigenvalue since, as we have seen, in this
case we have $|\alpha|=|\beta|=1/\sqrt{2}$. This type of degeneracy is to be
distinguished from the {\em nontrivial} degeneracy which occurs when eigenvectors
with different relative signs for the $z$'s are associated with the same eigenvalue.
As already noted, for $|\lambda|>2$ the relative signs of $z_{1}$, $z_{2}$, and $z_{3}$
in (7) are uniquely determined by $\lambda$ so that any eigenvector of the Bell operator
(2) violating inequality (1) (excepting the nondegenerate eigenvector corresponding
to the largest eigenvalue) is only {\em trivially} degenerate.

Turning to the $N$-particle case, one could equally prove that an $N$-particle state
giving the largest violation of Bell's inequality has to be maximally entangled. Properly
speaking, such an $N$-particle state will in general consist of a superposition of
maximally entangled states. That this requeriment has to be met for those states yielding
the {\em maximum} violation follows in a rather straightforward way when one considers
the correlation function \mbox{$P(\phi_{1},\, \phi_{2}, \ldots, \, \phi_{N};\, \Psi)$}
that quantum mechanics predicts for a general (pure) state of the form
\begin{equation}
\left| \Psi \right\rangle = \sum_{z_{1},\, z_{2}, \ldots,\, z_{N}}
a_{\,z_{1},\, z_{2}, \ldots,\, z_{N}} \left| \, z_{1},\, z_{2}, \ldots,\, z_{N} \right
\rangle \, , 
\end{equation}
with
\begin{displaymath}
a_{\,z_{1},\, z_{2}, \ldots,\, z_{N}} = \left| \, a_{\,z_{1},\, z_{2}, \ldots,\, z_{N}}
\right| e^{ {\rm i} \theta_{\scriptstyle z_{1},\, z_{2}, \ldots,\, z_{N}}} \, ,
\;\;\; \sum_{z_{1},\, z_{2}, \ldots,\, z_{N}} \left|\, a_{\,z_{1},\, z_{2}, \ldots,\,
z_{N}} \right| ^{2} = 1 \, ,
\end{displaymath}
and where $|z_{i}\rangle$ represents the eigenvector of the spin operator along the
$z$-axis for the {\em i\/}th particle $(i=1,\, 2,\ldots,\, N)$, with eigenvalue
$z_{i}=\pm1$. As before, and without loss of generality, we assume that each particle
is subjected to a spin measurement along a direction lying in the $x$--$y$ plane, with
azimuthal angle $\phi_{i}$. The quantum prediction for \mbox{$P(\phi_{1},\, \phi_{2},
\ldots, \, \phi_{N};\, \Psi)$} is given by
\begin{eqnarray}
P(\phi_{1},\, \phi_{2},\ldots, \, \phi_{N};\, \Psi) \!\!\!& = & 
\!\!\! 2\sum_{z_{1},\, z_{2}, \ldots,\, z_{N}} \!\!\!\!\!\!\!\! {^{\ast}} \;\;\:
\left|\, a_{\,z_{1},\, z_{2}, \ldots,\,
z_{N}} \right| \left|\, a_{\,-z_{1},\, -z_{2}, \ldots,\, -z_{N}} \right|
\nonumber \\
 & & \!\!\! \times \;\cos \left( \phi_{\,z_{1},\, z_{2}, \ldots,\, z_{N}} - 
z_{1}\phi_{1} - z_{2}\phi_{2} - \cdots - z_{N}\phi_{N} \right) \, , \nonumber \\
 & &
\end{eqnarray}
where $\phi_{\,z_{1},\, z_{2}, \ldots,\, z_{N}}=\theta_{\,-z_{1},\, -z_{2},
\ldots,\, -z_{N}} - \,\theta_{\,z_{1},\, z_{2}, \ldots,\, z_{N}}$, and where the
asterisk indicates that, for each pair of combinations of indices $z_{1},
\, z_{2}, \ldots,\, z_{N}$ and $-z_{1},\, -z_{2}, \ldots,\, -z_{N}$, the summation runs
over either the indices $z_{1},\, z_{2}, \ldots,$\linebreak $z_{N}$ or $-z_{1},\,
-z_{2}, \ldots, \, -z_{N}$. By examining Eq. (15), one finds that for the function
\mbox{$P(\phi_{1},\, \phi_{2}, \ldots, \, \phi_{N};\, \Psi)$} to take on its extreme values
$\pm$ it is necessary that $ \left| \, a_{\,z_{1},\, z_{2}, \ldots,\, z_{N}} \right| =
\left| \, a_{\,-z_{1},\, -z_{2}, \ldots,\, -z_{N}} \right| $ for all $ z_{1},\, z_{2},
\ldots,\, z_{N}$. This is because, if this condition is fulfilled, then we have
$ 2 \, \sum^{\,\ast}_{ z_{1},\, z_{2}, \ldots,\, z_{N}} 
\left|\, a_{\,z_{1},\, z_{2}, \ldots,\, z_{N}} \right|^{2} = 1 $, and thus expression
(15) may attain the value $\pm1$ for a suitable choice of the relative phases
$\phi_{\,z_{1},\, z_{2}, \ldots,\, z_{N}}$, namely, for $\phi_{\,z_{1},\, z_{2}, 
\ldots,\, z_{N}} = z_{1}\phi_{1} + z_{2}\phi_{2} + \cdots + z_{N}\phi_{N} + n\pi $,
where $n=0,\, \pm1,\, \pm2,\ldots$ . It thus follows that a state vector of the form
\begin{equation}
\left| \Psi \right\rangle = \sum_{z_{1},\, z_{2}, \ldots,\, z_{N}} \!\!\!\!\!\!\!\! 
{^{\ast}} \;\;\: c_{\,z_{1},\, z_{2}, \ldots,\, z_{N}} \left| \Psi (z_{1},\, z_{2}, 
\ldots,\, z_{N}) \right\rangle \, , 
\end{equation}
with $\sum^{\,\ast}_{ z_{1},\, z_{2}, \ldots,\, z_{N}}
\left|\, c_{\,z_{1},\, z_{2}, \ldots,\, z_{N}} \right|^{2} = 1$, and
\begin{eqnarray}
\left| \Psi(z_{1},\, z_{2}, \ldots,\, z_{N}) \right\rangle \!\!\! &  =  &
\!\!\!  \frac{1}{\sqrt{2}} \,( \, \left| \, z_{1},\, z_{2}, \ldots,\, z_{N} \right\rangle
\pm e^{ {\rm i}(z_{1}\phi_{1} + z_{2}\phi_{2} + \cdots + z_{N}\phi_{N}) }
 \nonumber   \\
& &  \;\;
  \times \, \left. \left| \, -z_{1},\, -z_{2}, \ldots,\, -z_{N} \right\rangle \,
\right) \, , 
\end{eqnarray}
will yield the value $\pm1$ for \mbox{$P(\phi_{1},\, \phi_{2}, \ldots, \, \phi_{N};\, 
\Psi)$}, and then it will be able to violate maximally a Bell inequality built up from
correlation functions involving spin measurements in the $x$--$y$ plane. State vector (13)
can be regarded as the simplest instance of Eq. (16). In this case only one term appears
because either of the eigenvectors associated with the extreme eigenvalues $\lambda=4$ or
$\lambda=-4$ of the Bell operator (12) is nondegenerate. In general the number of terms
appearing in expansion (16) will be equal to the dimensionality of the ({\em nontrivially})
degenerate subspace corresponding to the maximum eigenvalue of the relevant Bell operator.

The Bell operator \B\ for the Hardy inequality (1) takes {\samepage the general form 
\cite{Hardy}}
\pagebreak
\begin{eqnarray}
B_{{\rm H}} \!\!\!& = & \!\!\!\alpha\, \sigma\left (\phi_{1}^{\alpha} \right) \,
\sigma\left (\phi_{2}
^{\alpha}\right) \cdots \sigma\left (\phi_{N}^{\alpha}\right) + \beta\, \sigma\left
(\phi_{1}^{\beta}\right) \,  \sigma \left(\phi_{2}^{\beta}\right) \cdots \sigma\left
(\phi_{N}^{\beta}\right)  \nonumber   \\
& & + \, \gamma\, \sigma\left (\phi_{1}^{\gamma} \right) \, \sigma\left (\phi_{2}
^{\gamma}\right) \cdots \sigma\left (\phi_{N}^{\gamma}\right) + \delta\, \sigma\left
(\phi_{1}^{\delta}\right) \,  \sigma\left (\phi_{2}^{\delta}\right) \cdots \sigma\left
(\phi_{N}^{\delta}\right) \, , \nonumber \\
& &
\end{eqnarray}
where, as before, $\alpha$, $\beta$, $\gamma$, and $\delta$ are sign factors with
$\alpha\beta\gamma\delta =-1$. As was mentioned, the state vectors violating inequality
(1) will be those eigenvectors of (18) with eigenvalues $\lambda$ such that $2<|\lambda|
\leq 4$. Naturally, the four parameters $\phi_{i}^{\alpha}$, $\phi_{i}^{\beta}$,
$\phi_{i}^{\gamma}$, and $\phi_{i}^{\delta}$ are not all independent. Indeed, for each
value of $i$, there are the following possibilities \cite{Hardy}: (i) $\phi_{i}^{\beta}=
\phi_{i}^{\alpha}$, $\phi_{i}^{\delta} = \phi_{i}^{\gamma}$; (ii) $\phi_{i}^{\gamma}=
\phi_{i}^{\alpha}$, $\phi_{i}^{\delta} = \phi_{i}^{\beta}$; and (iii) $\phi_{i}^{\delta}=
\phi_{i}^{\alpha}$, $\phi_{i}^{\gamma} = \phi_{i}^{\beta}$. In any case the most general
eigenvector for the Bell operator (18) is one of the form (16) with \mbox{$\left| \Psi 
(z_{1},\, z_{2}, \ldots,\, z_{N}) \right\rangle$} given by
\begin{eqnarray}
\left| \Psi (z_{1},\, z_{2}, \ldots,\, z_{N})  \right\rangle  \,
= & \!\!\! \alpha_{\, z_{1},\, z_{2},\ldots,\, z_{N}} & \!\!\!\! \left| \, 
z_{1},\, z_{2}, \ldots,\, z_{N} \right\rangle \,\,\,\,   \nonumber  \\
\,+ \,\, \beta_{\, z_{1},\, z_{2},\ldots,\, z_{N}} & \!\!\!\!  e^{ {\rm i} \phi_{
\scriptstyle z_{1}, \, z_{2}, \ldots,\, z_{N}}} & \!\!\!\! \left| \, -z_{1},\,
-z_{2}, \ldots,\, -z_{N} \right\rangle \, ,
\end{eqnarray}
where $\alpha_{\, z_{1},\, z_{2},\ldots,\, z_{N}}$ and $\beta_{\, z_{1},\, z_{2},\ldots,
\, z_{N}}$ are real numbers with $\alpha^{2}_{ z_{1},\, z_{2},\ldots,\, z_{N}} + 
\beta^{2}_{ z_{1},\, z_{2},\ldots,\, z_{N}} = 1$. This is so because, as it stands,
the state vector (16) with \mbox{$\left| \Psi (z_{1},\, z_{2}, \ldots,\, z_{N}) \right
\rangle$} given by (19) turns out to be the most general (pure) state for $N$ spin-1/2
particles and then it will be always possible for {\em any} eigenvector of the Bell
operator (18) to be arranged so as to fit in with the form displayed by such Eqs. (16)
and (19). Also generally, for any eigenvalue $\lambda$ the summation in (16) will extend
over all those ({\em nontrivially}) degenerate eigenvectors (19) associated with that given
eigenvalue. Of course, the coefficients $\alpha_{\, z_{1},\, z_{2},\ldots,\, z_{N}}$ and
$\beta_{\, z_{1},\, z_{2},\ldots, \, z_{N}}$ (or, rather, their product), as well as the
phase factors $\phi_{z_{1}, \, z_{2}, \ldots,\, z_{N}}$ will depend on the actual value of
$\lambda$. So, for the case in which $|\lambda|$ attains its maximum value 4, (i.e., when
the inequality is maximally violated), we must have $\phi_{z_{1}, \, z_{2}, \ldots,\,
z_{N}} = z_{1}\phi_{1}+z_{2}\phi_{2}+\cdots+z_{N}\phi_{N}+n\pi$, and
$|\alpha_{\, z_{1},\, z_{2},\ldots,\, z_{N}}| = |\beta_{\, z_{1},\, z_{2},\ldots,\, z_{N}}|
= 1/{\sqrt{2}}$, for all $z_{1},\, z_{2},\ldots,\, z_{N}$ (see Eq. (17)). It is easy to
show, however, that this latter requirement should be fulfilled not only by the maximum
possible eigenvalue of the relevant Bell operator but also by its {\em largest} one. For
this purpose, let us consider the expectation value of the operator (18) for the state
vector (16) with \mbox{$\left| \Psi (z_{1},\, z_{2}, \ldots,\, z_{N}) \right\rangle$}
given by (19). This expectation value is
\begin{eqnarray}
\!\!\!\!\!  \left \langle \Psi \left| B_{{\rm H}} \right| \Psi \right\rangle 
\!\!\! &  = \!\!\!\! &
\sum_{z_{1},\, z_{2}, \ldots,\, z_{N}} \!\!\!\!\!\!\!\! {^{\ast}} \;\;\:
2\, \alpha_{\,z_{1},\, z_{2}, \ldots,\, z_{N}} \,\beta_{\,z_{1},\, z_{2}, \ldots,\, z_{N}}
\left|\, c_{\,z_{1},\, z_{2}, \ldots,\, z_{N}} \right|^{2}  \nonumber   \\
& &  \,\,\, \times \left[ \,A_{\alpha} \left( z_{1},\, z_{2},\ldots,\, z_{N} \right)
+ B_{\beta} \left( z_{1},\, z_{2},\ldots,\, z_{N} \right)  \right.   \nonumber    \\
& & \,\,\,\,\,\,\,\,  +  \left.  C_{\gamma} \left( z_{1},\, z_{2},\ldots,\, z_{N} \right)
+  D_{\delta} \left( z_{1},\, z_{2},\ldots,\, z_{N} \right) \, \right] \, ,
\end{eqnarray}
where, for example, $C_{\gamma} = \gamma \cos ( \phi_{\,z_{1},\, z_{2},
\ldots,\, z_{N}} -  z_{1}\phi_{1}^{\gamma} - z_{2}\phi_{2}^{\gamma} - \cdots -
z_{N}\phi_{N}^{\gamma} )$. Clearly, as the product \mbox{$\alpha_{\, z_{1},\, z_{2},
\ldots,\, z_{N}}\, \beta_{\, z_{1},\, z_{2},\ldots, \, z_{N}}$} factorises out in each
of the terms involved in Eq. (20), it follows that the largest eigenvalue for \B\ must
fulfil $|\alpha_{\, z_{1},\, z_{2},\ldots,\, z_{N}}| = |\beta_{\, z_{1},\, z_{2},
\ldots,\, z_{N}}| = 1/{\sqrt{2}}$, for all $z_{1},\, z_{2},\ldots,\, z_{N}$. In view
of Eq. (20) this conclusion holds irrespective of the number of correlation functions
$(A_{\alpha},\, B_{\beta},\, C_{\gamma},\, D_{\delta},\ldots\,)$ involved. Furthermore,
the structure of Eqs. (19) and (20) remain unchanged for the general case in which the
spin measurements are carried out along arbitrary directions (in fact, for this case,
we have only to redefine the meaning of the $|z_{i}\rangle$'s in (19) as denoting
states of spin-up $(z_{i}=+1)$ or -down $(z_{i}=-1)$ for particle $i$ along some
appropriate $z$-axis which, in general, will differ from one particle to the other).
Thus, we have demonstrated that a state of $N$ spin-1/2 particles $(N \geq 3)$ giving the
largest violation of a Bell inequality must generally consist of a superposition of
maximally entangled states. This conclusion applies in particular to those states
giving the maximum possible violation. This is achieved when each of the correlation
functions attains an appropriate extremum value $\pm1$. So, as a direct (``all or
nothing") contradiction with local realism arises just at the level of perfect
correlations, it follows that {\em any} state leading to such a contradiction should
in general involve a superposition of maximally entangled states. In fact, as we have
seen, any state vector yielding the value $\pm1$ for the correlation function
\mbox{$P(\phi_{1},\, \phi_{2}, \ldots, \, \phi_{N};\, \Psi)$} must necessarily
assume the form of Eqs. (16) and (17).

We conclude by noting that, as expected, this function factorises for a general product
state of the form $|\Psi\rangle = |\Psi_{1}\rangle \otimes |\Psi_{2}\rangle \otimes
\cdots \otimes |\Psi_{N}\rangle$, with $|\Psi_{i}\rangle = a_{\,z_{i}} |\,z_{i}\rangle + 
a_{\, -z_{i}} |-z_{i}\rangle$, $a_{\,z_{i}}=|a_{\,z_{i}}|\,e^{{\rm i}\theta_{\scriptstyle
z_{i}}}$, and $|a_{\,z_{i}}|^{2} + |a_{\,-z_{i}}|^{2} = 1$, $i=1,\, 2,\ldots,\, N$.
Indeed, by making use of the identity
\begin{equation}
2\sum_{z_{1},\, z_{2}, \ldots,\, z_{N}} \!\!\!\!\!\!\!\! {^{\ast}} \;\;\: \cos
\left( z_{1}\gamma_{1}+z_{2}\gamma_{2}+\cdots+z_{N}\gamma_{N} \right) =
2^{N} \cos\gamma_{1} \, \cos\gamma_{2} \ldots \cos\gamma_{N} \, ,
\end{equation}
one can readily see that expression (15) takes the form
\begin{equation}
P(\phi_{1},\, \phi_{2}, \ldots, \, \phi_{N};\, \Psi) = 2^{N} a_{0} \,
\cos(\phi_{1}+\eta_{1})\, \cos(\phi_{2}+\eta_{2}) \ldots \cos(\phi_{N}+\eta_{N})\, ,
\end{equation}
where $\eta_{i}=\theta_{z_{i}}-\, \theta_{-z_{i}}$, and $a_{0}$ is a constant with value
$a_{0}= |a_{\,z_{1}}||a_{\,z_{2}}| \cdots |a_{\,z_{N}}|$\linebreak
$\times |a_{\,-z_{1}}||a_{\,-z_{2}}| \cdots |a_{\,-z_{N}}|$. Note that $a_{0} 
\leq 2^{-N}$ and then, as it should be, $|P| \leq 1$. Obviously, $a_{0}$ reaches its
maximum value whenever $|a_{\,z_{i}}| = |a_{\,-z_{i}}|= 1/{\sqrt{2}}$, for all $i$.
In any case, it is apparent from (22) that for a product state the outcome of a spin
measurement for any one of the particles becomes completely uncorrelated with respect
to the outcomes corresponding to the other particles, and then such a state will be
unable to yield a violation of Bell's inequality.

\vspace{4mm}
{\bf Acknowledgment:} The author would like to acknowledge informative discussions
with Gregorio Rentero on the subject of quantum nonlocality.

\end{document}